\documentclass{iopart}

\usepackage{iopams}
\usepackage{epsf}
\usepackage{cite}

\newcommand{\bd}{\begin{displaymath}}
\newcommand{\ed}{\end{displaymath}}
\newcommand{\be}{\begin{equation}}
\newcommand{\ee}{\end{equation}}
\newcommand{\ba}{\begin{eqnarray}}
\newcommand{\ea}{\end{eqnarray}}

\begin{document}

\title{On Wheeler's delayed-choice {\it Gedankenexperiment} and its
laboratory realization}

\author{M Bo\v zi\'c$^1$, L Vu\v skovi\'c$^2$, M Davidovi\'c$^3$
and A S Sanz$^4$}
\address{$^1$Institute of Physics, University of Belgrade, 11080 Belgrade,
Serbia}
\address{$^2$Department of Physics, Old Dominion University, Norfolk,
Virginia, USA}
\address{$^3$Faculty of Civil Engineering, University of Belgrade,
11000 Belgrade, Serbia}
\address{$^4$Instituto de F\'{\i}sica Fundamental - CSIC, Serrano 123,
28006 - Madrid, Spain}

\eads{\mailto{bozic@ipb.ac.rs},\mailto{vuskovic@odu.edu},
\mailto{milena@grf.rs},\mailto{asanz@iff.csic.es}}

\begin{abstract}
Here, we present an analysis and interpretation of the experiment
performed by Jacques {\it et al.}\ (2007 {\it Science} {\bf 315} 966),
which represents a realization of Wheeler's delayed-choice
{\it Gedankenexperiment}.
Our analysis is based on the evolution of the photon state, since the
photon enters into the Mach-Zehnder interferometer with a removable
beam-splitter until it exits.
Given the same incident photon state onto the output beam-splitter,
$BS_{output}$, the photon's state at the exit will be very different
depending on whether $BS_{output}$ is on or off.
Hence, the statistics of photon counts collected by the two
detectors, positioned along orthogonal directions at the exit of the
interferometer, is also going to be very different in either case.
Therefore, it is not that the choice of inserting (on) or removing
(off) a beam-splitter leads to a delayed influence on the photon
behavior before arriving at the beam-splitter, but that such a choice
influences the photon state at and after $BS_{output}$, i.e., after it
has exited from the Mach-Zehnder interferometer.
The random on/off choice at $BS_{output}$ has no delayed effect on the
photon to behave as a wave or a corpuscle at the entrance and inside
the interferometer, but influences the subsequent evolution of the
photon state incident onto $BS_{output}$.
\end{abstract}






\section{Introduction}
\label{sec1}

Since the inception of Quantum Mechanics, various
{\it Gedankenexperimente} were proposed, which made evident properties
very different to those described by Classical Mechanics.
With time, the necessity to test these fundamental properties led, in
many cases, to the development of the technology necessary to pass from
mere ideas to real experiments with real particles ---either massive
particles or photons.

\begin{figure}
 \begin{center}
 \epsfxsize=7cm {\epsfbox{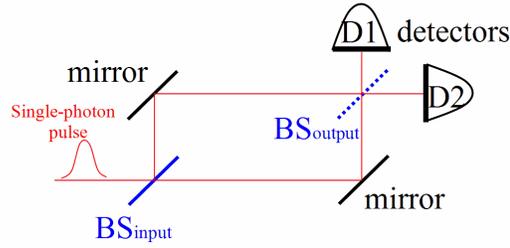}}
 \caption{\label{fig1} Schematics of an MZI with a removable
  beam-splitter at the output, as considered by Wheeler \cite{r1}.}
 \end{center}
\end{figure}

One of such experiments is the well-known Wheeler's delayed-choice
{\it Gedankenexperiment} \cite{r1}, proposed to test the nature of
wave or corpuscle of quantum particles by means of a Mach-Zehnder
interferometer (MZI).
In order to select one or the other behavior, the interferometer has a
removable (output) beam-splitter at the exit (see Fig.~\ref{fig1}).
When the output beam-splitter is positioned on its place, the MZI
configuration is said to be {\it closed}; when it is off place, the
configuration is {\it open}.
Following Wheeler's argument, with the first configuration one observes
the wave behavior of the particle, while with the latter, the
corpuscular one.

In 2007, Jacques {\it et al.}\ \cite{r2} carried out in the lab
Wheeler's experiment.
In this experiment, the choice between the open and closed
configurations was realized by means of an electro-optical modulator,
which could be switched at will between the two different
configurations in times of the order of 40~ns.
This time is enough to close or to open the MZI configuration once the
photon is inside.
Furthermore, with this distance in time, the switching of the output
beam-splitter and the entry of the photon into the MZI are events well
separated in time relativistically.
It is crucial there is no correlation in time between both events,
which have to take place once the photon is inside the interferometer,
as argued by Wheeler \cite{r1}.
Otherwise the photon might acquire some ``hidden information'' on the
chosen experimental configuration and could readjust its behavior
accordingly.
This was precisely the main reason leading Wheeler to formulate his
experiment against other alternative experiments proposed at the time
to test complementarity.

\section{Laboratory realization and looking backward interpretation}
\label{sec2}

In the experiment \cite{r2}, single photons are sent towards a 48-m
polarization interferometer, equivalent to a time-of-flight of about
160~ns.
A binary random number, 0 or 1, generated by a quantum random number
generator (QRNG), drives the electro-optical-modulator (EOM) voltage
between $V = 0$ and $V = V_\pi$ within 40~ns, after an electronic
delay of 80~ns.
Two synchronized signals from a clock are used to trigger the
single-photon emission and the QNRG.
Thus, the random choice between the open and closed MZI configurations
takes place when the photon is approximately about the central part of
the interferometer, long after it passed through $BS_{output}$.
Moreover, a phase shift, $\phi$, between the two MZI arms is introduced
by executing a tilt with a $BS_{output}$ piezoelectric actuator (PZT).

The chosen configuration, the detection events (which detector
registered the event), and the PZT position were then recorded for
each photon.
All raw data were saved in the real time.
For each PZT-position (phase), detection events at $D1$ and $D2$
corresponding to each configuration were sorted out.
Thus, when analyzing these data, one observes \cite{r2}:
\begin{itemize}
 \item[A)] The counts at $D1$ and $D2$ display sinusoidal oscillations
 with for the closed configuration (see Fig.~3A in Ref.~\cite{r2}).

 \item[B)] The counts at $D1$ and $D2$ do not depend on $\phi$ for the
 open configuration (see Fig.~3B in Ref.~\cite{r2}).
\end{itemize}
From these experimental results, Jacques {\it et al.}\ concluded
\cite{r2} that ``{\it the behavior of the photon at the first beam
splitter depends on the choice of the observable that is measured
behind the output beam splitter, even when that choice is made at a
position and a time such that it is separated from the entrance of the
photon into the interferometer by a space-like interval}'', which
according to Wheeler \cite{r1} translates as ``{\it a strange inversion
of the normal order of time.}''


\section{Looking forward interpretation of the experiment}
\label{sec3}

Here, we analyze and propose an the intuitive interpretation of the
experiment of Jacques {\it et al.}\ \cite{r2} by considering the
evolution of the photon state from its entrance into the MZI and
throughout its passage.

Beam-splitters are essential constituting elements of a MZI.
In considering the action of a beam-splitter on a quantum object, it is
fundamental to take into account the incident quantum state of such an
object as well as the subsequent evolution of this state
\cite{r3,r4,r5,r6,r7,r8}.
In this sense, a beam-splitter can be considered as a transformer of
an incident wave field (photon field or matter wave field) into a field
which has narrow maxima at the points along and in close vicinity of
two or several specific directions.
This becomes evident when one considers a thin grating as a model
for a beam splitter for photons \cite{r3,r4,r5}, atoms and molecules
\cite{r6}.
From such considerations, it follows that a lossless beam-splitter can
also be termed as a {\it coherent beam-splitter} \cite{r7}, since the
outgoing ``separated beams'' do not spread independently, but jointly,
keeping their mutual coherence.

Taking this into account, we note the following:
\begin{itemize}
 \item[i)] The input and output beam-splitters (in the latter case,
 when it is on) transform two different states of a photon.
 Therefore, the probabilities associated with the photon going through
 one or the other characteristic direction (of the two possible) behind
 $BS_{input}$ and $BS_{output}$ are different.

 \item[ii)] The state of the photon incident onto $BS_{output}$ is
 determined by the interaction with $BS_{input}$, the two mirrors and
 the free-evolution equation.
 Therefore, the photon state incident onto $BS_{output}$ is independent
 of whether $BS_{output}$ is on or off.

 \item[iii)] The evolution of a given photon state incident onto
 $BS_{output}$ depends on whether $BS_{output}$ is on or off.
 Therefore, the photon state at the exit of $BS_{output}$ when the
 latter is on is very different from the photon state when this
 beam-splitter is off.
 The statistics measured by the detectors $D1$ and $D2$ when
 $BS_{output}$ is off is then different from the statistics when it
 is on, because the state of the outgoing photon depends on the on/off
 state of $BS_{output}$.
\end{itemize}

When $BS_{output}$ is on, it changes the incidence photon state, which
then evolves freely outside the interferometer.
The probability that a photon chooses one or the other direction is
determined by its incident state and the interaction with a grating.
When the beam-splitter is off, the incidence photon state (as well
as the photon itself) propagates freely. Consequently, a photon keeps
moving through the direction along which it was moving before reaching
$BS_{output}$.

The on/off switching of $BS_{output}$ does not influence the behavior
of the photon before it has arrived to this beam-splitter.
Such a switching influences the photon state both at $BS_{output}$ and
at the exit from this beam-splitter.
Therefore, the photon statistics at the detectors will also be
influenced by the switching.

So, we conclude that there is no delayed-choice action.
The on/off switching does not decide whether the photon will move
along one or both routes after it has already completed its travel;
the on/off switching does influence the evolution of the photon state
incident onto $BS_{output}$.
In our opinion, the experiment of Jacques {\it et al.}\ \cite{r2}
proves that the wave and corpuscle properties of photons are
compatible, i.e., both are present in the same experiment.


\section{Mode operators and photon statistics at the exit of the MZI in
the open and closed configurations}
\label{sec4}

\begin{figure}
 \begin{center}
 \epsfxsize=4cm {\epsfbox{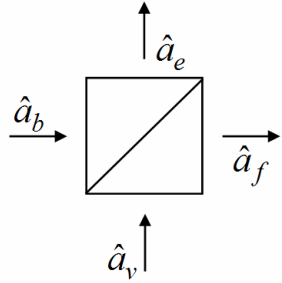}}
 \caption{\label{fig2} Schematic representation of the modes describing
 a beam-splitter.}
 \end{center}
\end{figure}

\begin{figure}
 \begin{center}
 \epsfxsize=7cm {\epsfbox{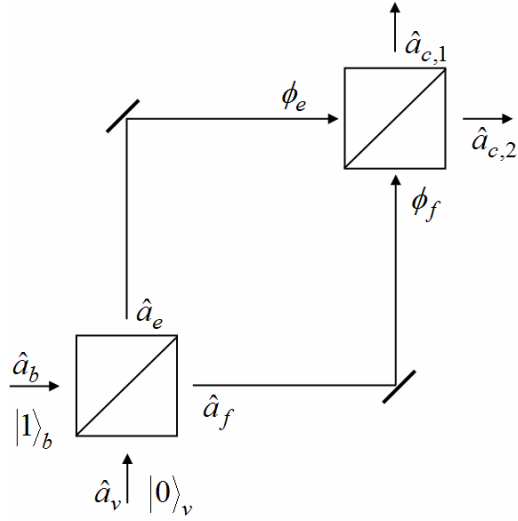}}
 \caption{\label{fig3} Schematic representation of the modes describing
 the MZI closed configuration, where $BS_{output}$ is on.}
 \end{center}
\end{figure}

The above conclusions can be alternatively derived in an elegant
manner by considering a second-quantization treatment of the photon
electromagnetic field in the MZI.
To do so, note that the relationship between the input and output modes
of the electromagnetic field surrounding a beam-splitter (see
Fig.~\ref{fig2}) are now well known \cite{r8}:
\be
 \hat{a}_s = R \hat{a}_b + T \hat{a}_v , \qquad
 \hat{a}_f = T \hat{a}_b + R \hat{a}_v ,
 \label{e1}
\ee
The complex transmission and reflection coefficients for a lossless
beam-splitter satisfy the relations
\be
 \begin{array}{l}
    RT^* + TR^*  = 0 , \\
   |R|^2 + |T|^2 = 1 .
 \end{array}
 \label{e2}
\ee
As can be seen in Fig.~\ref{fig3}, the MZI closed configuration
consists of two beam-splitters and two mirrors.
The second beam-splitter may be tilted in order to introduce a phase
shift.
Thus, three sets of modes are necessary to describe the photon states
in the MZI, namely $(\hat{a}_b,\hat{a}_v)$, $(\hat{a}_e,\hat{a}_f)$ and
$(\hat{a}_{c,1},\hat{a}_{c,2})$.

\begin{figure}
 \begin{center}
 \epsfxsize=7cm {\epsfbox{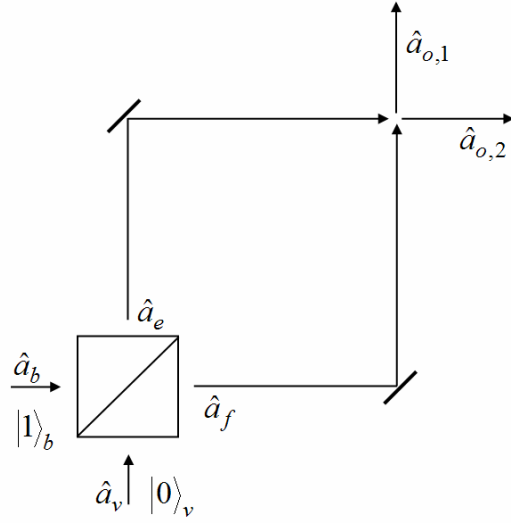}}
 \caption{\label{fig4} Schematic representation of the modes describing
 the MZI open configuration, where $BS_{output}$ is off.}
 \end{center}
\end{figure}

The relations between the output modes, $(\hat{a}_{c,1},\hat{a}_{c,2})$,
and the internal ones, $(\hat{a}_e,\hat{a}_f)$, are similar to relations
(\ref{e1}), but containing the additional phase shifts $\phi_e$ and
$\phi_f$ in order to account for the tilting of $BS_{output}$,
\be
 \begin{array}{l}
  \hat{a}_{c,1} = R \hat{a}_e e^{i\phi_e}
   + T \hat{a}_f e^{i\phi_f} , \\
  \hat{a}_{c,2} = R \hat{a}_f e^{i\phi_f}
   + T \hat{a}_e e^{i\phi_e} .
 \end{array}
 \label{e3}
\ee
Now, using relations (\ref{e1}) one can determine the relationship
between the output and input modes, which reads as \cite{r8}
\be
 \begin{array}{l}
  \hat{a}_{c,1} = R_{MZ} \hat{a}_b + T_{MZ} \hat{a}_v , \\
  \hat{a}_{c,2} = T_{MZ} \hat{a}_b + R'_{MZ} \hat{a}_v ,
 \end{array}
 \label{e4}
\ee
where
\be
 \begin{array}{l}
  R_{MZ}  = R^2 e^{i\phi_e} + T^2 e^{i\phi_f} , \\
  R'_{MZ} = R^2 e^{i\phi_f} + T^2 e^{i\phi_e} , \\
  T_{MZ}  = RT \left( e^{i\phi_e} + e^{i\phi_f} \right) .
 \end{array}
 \label{e5}
\ee
In the open configuration of the MZI (see Fig.~\ref{fig4}), input and
internal modes are the same as in the closed configuration:
$(\hat{a}_b,\hat{a}_v)$, $(\hat{a}_e,\hat{a}_f)$.
We shall denote the output modes by $(\hat{a}_{o,1},\hat{a}_{o,2})$.
Since the output beam-splitter is off, the relations between output and
input modes in the MZI open configuration are:
\be
 \hat{a}_{o,1} = T \hat{a}_b + R \hat{a}_v , \qquad
 \hat{a}_{o,2} = R \hat{a}_b + T \hat{a}_v ,
 \label{e6}
\ee
Determining now the photon statistics behind the MZI in the open and
closed configurations is straightforward.
In the open configuration, the mean photon numbers at the detectors
will be
\be
 \begin{array}{l}
  \displaystyle \frac{N_{o,1}}{N} =
   \langle \hat{n}_{o,1} \rangle =
   {_v\langle} 0| {_b\langle} 1| \hat{a}_{o,1}^+ \hat{a}_{o,1}
    |1\rangle_b |0\rangle_v = |T|^2 , \\
  \displaystyle \frac{N_{o,2}}{N} =
   \langle \hat{n}_{o,2} \rangle =
   {_v\langle} 0| {_b\langle} 1| \hat{a}_{o,2}^+ \hat{a}_{o,2}
    |1\rangle_b |0\rangle_v = |R|^2 , \\
 \end{array}
 \label{e7}
\ee
with $N$ being the total number of incident photons.
Therefore, the numbers of photons that propagate towards detectors $D1$
and $D2$ do not depend on the phase $\phi$, which is in agreement with
the experiment.
This is simple to understand.
If the beam-splitter is off, there is no way its tilt can influence the
motion and passage state of a photon.
On the other hand, in the closed configuration, the number of photons
at the detectors will be
\be
 \begin{array}{l}
  \displaystyle \frac{N_{c,1}}{N} =
   \langle \hat{n}_{c,1} \rangle =
   {_v\langle} 0| {_b\langle} 1| \hat{a}_{c,1}^+ \hat{a}_{c,1}
    |1\rangle_b |0\rangle_v = |R_{MZ}|^2 \\
    \qquad = |R|^4 + |T|^4 - 2 |R|^2 |T|^2 \cos \phi , \\
  \displaystyle \frac{N_{c,2}}{N} =
   \langle \hat{n}_{c,2} \rangle =
   {_v\langle} 0| {_b\langle} 1| \hat{a}_{c,2}^+ \hat{a}_{c,2}
    |1\rangle_b |0\rangle_v = |T_{MZ}|^2 \\
    \qquad = 2 |R|^2 |T|^2 \left( 1 - \cos \phi \right) ,
 \end{array}
 \label{e8}
\ee
where
\be
 \phi \equiv \phi_e - \phi_f .
 \label{e9}
\ee
In deriving Eq.~\ref{e8}, the moduli and phases of the complex
coefficients $R$ and $T$ were introduced taking into account relations
\ref{e2},
\be
 R = |R| e^{i\varphi_R} , \qquad T = |T| e^{i\varphi_T} , \qquad
 \varphi_R - \varphi_T = \frac{\pi}{2} ,
 \label{e10}
\ee
Assuming $|R| = |T| = 1/\sqrt{2}$, one finds simpler relations for the
number of photons along the two directions in the closed configuration:
\be
 \begin{array}{l}
  \displaystyle N_{c,1} = \frac{N}{2}\ \left( 1 - \cos \phi \right) , \\
  \displaystyle N_{c,2} = \frac{N}{2}\ \left( 1 + \cos \phi \right) .
 \end{array}
\ee

\section{Comparison of the arguments leading to the two different
interpretations}
\label{sec5}

The reasoning leading to the conclusion that particle properties are
complementary \cite{r1,r2} and that ``we have a strange inversion of
the normal order of time'' \cite{r1,r2} were based on the following
two statements:
\begin{itemize}
 \item[1)] When $BS_{output}$ is off, the number of detected photons
 $N_{0,1}$ at the detector $D1$ is equal to the number of detected
 photons $N_{0,2}$ at $D2$.
 In this case, we have $N_{0,1} = N_{0,2} = N/2$, which does not depend
 on the phase $\phi$.
 Hence, one measures the corpuscle property associated with the photon.

 \item[2)] When $BS_{output}$ is on, the number of photons detected,
 $N_{c,1}$ and $N_{c,2}$, depend on the phase $\phi$, i.e., $N_{c,1}
 = N_{c,1}(\phi)$ and $N_{c,2} = N_{c,2}(\phi)$.
 In this case, therefore, one measures a wave property of the photon.
\end{itemize}
This reasoning is based on the statement that the measurement of the
same quantity, namely number of photons, some times has the meaning of
a particle property measurement and other times it acquires the meaning
of a wave property measurement.

On the contrary, the reasoning leading to the conclusion that particle
and wave properties are compatible is based on the statement that the
evolution of the same photon state incident onto the output beam
splitter depends on whether this beam splitter is on or off.
As a consequence, the relations between creation and annihilation
operators associated with the input and output modes are different
in the on and off cases:
\begin{itemize}
 \item[1)] When the beam splitter is off, the wave function of each
 single photon incident onto $BS_{output}$ evolves freely.
 Because of that, the number of photons at the detectors is determined
 by (7), i.e., it does not depend on the phase.
 The numbers at the detectors are equal to the numbers of photons
 arriving from the corresponding directions to $BS_{output}$.

 \item[2)] When $BS_{output}$ is on, it influences the wave function
 evolution of each arriving photon.
 Consequently, the number of photons moving at the exit towards one or
 the other detector is changed, depending of the specific property of
 $BS_{output}$, e.g., its tilt, which reflects in the phase.
\end{itemize}

The reasoning that takes into account the wave function of each photon
for both cases, on and off, leads to the  consistent explanation of why
the number of photons at the detectors is constant when $BS_{output}$
is off and varies with the tilt when it is on.
It means that wave and particle properties of the photon are present
simultaneously in the open and closed configurations.
Thus, the time-ordering of events arises as a natural consequence within
this reasoning, i.e., there is no ``strange inversion of the normal
order of time''.


\ack

M. Bo\v zi\'c and M. Davidovi\'c acknowledge support from the Ministry
of Science of Serbia through projects III45016 and OI171005.
A. S. Sanz acknowledges the Ministerio de Ciencia e Innovaci\'{o}n (Spain)
for support through projects FIS2007-62006 and FIS2010-22082 as well
as for a ``Ram\'on y Cajal'' Research Contract.


\end{document}